\documentclass[a4paper,UKenglish,cleveref, autoref]{lipics-v2016}
\usepackage{xcolor}
\usepackage{etoolbox}

\title{Formal verification of trading in financial markets} 

\author[1]{Suneel Sarswat}
\author[2]{Abhishek Kr Singh}
\affil[1]{Tata Institute of Fundamental Research, Mumbai, India\\
	\texttt{suneel.sarswat@gmail.com}}
\affil[2]{Tata Institute of Fundamental Research, Mumbai, India\\
	\texttt{abhishek.uor@gmail.com}}
\authorrunning{Suneel Sarswat and A.\,K. Singh} 


\EventEditors{}
\EventNoEds{2}
\EventLongTitle{}
\EventShortTitle{}
\EventAcronym{}
\EventYear{2019}
\EventDate{}
\EventLocation{}
\EventLogo{}
\SeriesVolume{}
\ArticleNo{}

\makeatletter
\patchcmd{\@verbatim}
  {\verbatim@font}
  {\verbatim@font\small}
  {}{}
\makeatother

\begin{document}

\newcommand{\tw}{\texttt}

\maketitle

\begin{abstract}

We introduce a formal framework for analyzing trades in financial markets. 
An exchange is where multiple buyers and sellers participate to trade.
These days, all big exchanges use computer algorithms that implement double sided auctions to match buy and sell requests and these algorithms must 
abide by certain regulatory guidelines.
For example, market regulators enforce that a matching produced by exchanges should 
be \emph{fair}, \emph{uniform} and \emph{individual rational}. 
To verify these properties of trades, we first formally define these notions
in a theorem prover and then give formal proofs of relevant results on matchings.
Finally, we use this framework to verify properties of two important classes of double
sided auctions. All the definitions and results presented in this paper are completely formalized in
the Coq proof assistant without adding any additional axioms to it.

\end{abstract}

\section{Introduction}
\label{section1}

In this paper, we introduce a formal framework for analyzing trades in financial markets. Trading is a principal component of all modern economies. Over the past few centuries, more and more complex instruments are being introduced for trade in the financial markets. All big stock exchanges use computer algorithms to match buy requests (demands) with sell requests (supplies) of traders. Computer algorithms are also used by  traders to place orders in the markets.\footnote{This is known as algorithmic trading.} With the arrival of computer assisted trading, the volume and liquidity in the markets have increased drastically and as a result, the markets have become more complex.

Software programs that enable the whole trading process are extremely complex and have to meet high efficiency criterion as they operate on massive amounts of data and in real time. Furthermore, to increase the confidence of traders in the markets, the market regulators set stringent safety and fairness guidelines for these software. Traditionally, to meet such criteria, software development has extensively relied on testing the programs on large data sets. Although testing is helpful in identifying bugs, it cannot guarantee the absence of bugs. Even small bugs in the trading software can have a catastrophic effect on the overall economy. An adversary might exploit a bug to his benefit and to the disadvantage of other genuine traders. These events are certainly undesirable in a healthy economy. 

Recently, there have been various instances \cite{nyse1, nyse2, nse} of violation of the trading rules by the stock exchanges. For example, in \cite{nyse1}, a regulator noted: "NYSE Arca failed to execute a certain type of limit order under specified market conditions despite having a rule in effect that stated that NYSE Arca would execute such orders"\footnote{The New York Stock Exchange and the Archipelago Exchange merged together to form NYSE Arca, which is an exchange where both stocks and options are traded.}. This is an instance of a program not meeting its specification. Here the program is a matching algorithm used by the exchange and the regulatory guidelines are the broad specifications for the program. Note that, in most of the cases, the guidelines stated by the regulators are not a complete specification of the program. Moreover, there is no formal guarantee that these  guidelines are consistent. These are some serious issues potentially compromising the safety and integrity of the markets. 

Recent advances in formal methods in computer science can be put to good use in ensuring safe and fair financial markets. During the last few decades, formal method tools have been increasingly successful in proving the correctness of large software and hardware systems \cite{seqcir,blast,compcert,sel4}. While Model checking tools have been used for the verification of hardware, the use of Interactive theorem provers have been quite successful in the verification of large software. A formal verification of financial algorithms using these tools can be helpful in the rigorous analysis of market behaviour at large. The matching algorithms used by the exchanges (venues) are at the core of the broad spectrum of algorithms used in financial markets. A formal framework for verifying matching algorithms can also be useful in verifying other algorithms in financial markets. 

\subsection{An overview of trading at an exchange}

An exchange is an organised financial market. There are various types of exchanges: stock exchange, commodity exchange, foreign exchange etc. An exchange facilitates trading between buyers and sellers for the products which are registered at the exchange. A potential trader (buyer or seller) places orders in the markets for a certain product. These orders are matched by the stock exchange to execute trades.  Most stock exchanges hold trading in two main sessions: pre-market (or auction session) and continuous  market (or regular trading session). 

The pre-market session reduces uncertainty and volatility in the market by discovering the opening price of the product. During the pre-market session, an exchange collects all the buy requests (bids) and sell requests (asks) for a fixed duration. At the end of this duration the exchange matches these buy and sell requests at a single price using a matching algorithm. In the continuous market session, the incoming buyers and sellers are continuously matched to each other. An incoming bid (ask), if matchable, is immediately matched to the existing asks (bids). Otherwise, if the bid (ask) is not matchable, it is placed in a priority queue prioritised by price. A trader can place orders of multiple quantity of each product to trade during both the sessions. However, for the analysis of the markets, it suffices to assume that each order is of a single unit of a single product; a multiple quantity order can always be treated as a bunch of orders each with a single quantity and the analysis for a single product will apply for all the products individually. As a result, note that a single trader who places an order of multiple units is seen as multiple traders ordering a single unit each, even in the continuous market. Thus, in both sessions of trade, multiple buyers and sellers are matched simultaneously. A mechanism used to match multiple buyers and sellers is known as the double sided auction \cite{friedman}.

In double sided auctions, an auctioneer (e.g. exchanges) collects buy and sell requests over a period of time. Each potential trader places the orders with a limit price: below which a seller will not sell and above which a buyer will not buy. The exchange at the end of this time period matches these orders based on their limit prices. This entire process is completed using a double sided auction matching algorithm. Designing algorithms for double sided auctions is well studied \cite{mcafee1992, WurmanWW98,NiuP13}. A major emphasis of many of these studies have been to either maximize the number of matches or  maximize the profit of the auctioneer. In the auction theory literature, the profit of an auctioneer is defined as the difference between the limit prices of matched bid-ask pair. However, most exchanges today earn their profit by charging transaction costs to the traders. Therefore, maximizing the number of matches increases the profit of the exchange as well as the liquidity in the markets.  There are other important properties, like fairness, uniformity and individual rationality, besides the number of matches which are considered while evaluating the effectiveness of a matching algorithm. However, it is known that no single algorithm can possess all of these properties simultaneously \cite{WurmanWW98,mcafee1992}. 

\subsection{Related work}\label{sec:related}
There is no prior work known to us which formalizes financial algorithms used by the exchanges. Passmore and Ignatovich in \cite{PassmoreI17} highlight the significance, opportunities and challenges involved in formalizing financial markets. Their work describes in detail the whole spectrum of financial algorithms that need to be verified for ensuring safe and fair markets. Matching algorithms used by the exchanges are at the core of this whole spectrum.  

On the other hand, there are quite a few works  formalizing various concepts from auction theory \cite{toolbox,lange,frank}. Most of these works focus on the Vickrey auction mechanism. In Vickrey auction, there is a single seller with different items and multiple buyers with valuations for every subsets of items. Each buyer places bids for every combination of the items. At the end of bidding, the aim of seller is to maximise total value of the items by suitably assigning the items to the buyers.

\subsection{Our contribution}
In this work, we formally define various notions from auction theory relevant for the analysis of trades in financial markets. We define notions like bids, asks and matching in the Coq proof assistant. The dependent types of Coq turn out to be very useful in giving concise representation to these notions, which also reflects their natural definitions. After preparing the basic framework, we define important properties of matching in a double sided auction: fairness, uniformity, maximality and individual rationality. These properties reflect various regulatory guidelines for trading. Furthermore, we formally prove some results on the existence of various combinations of these properties. For example, fairness and maximality can always be achieved simultaneously. These results can also be interpreted as consistency proofs for various subsets of regulatory guidelines. We prove all these results in the constructive setting of the Coq proof assistant without adding any additional axioms to it. These proofs are completed using computable functions which computes the actual instances (certificate). We also use computable functions to represent various predicates on lists. Finally, we use this setting to verify properties of two important classes of matching algorithms: dynamic price and uniform price algorithms. 

In Section~\ref{sec:modeling}, we formally define the theory of double sided auctions. In Section~\ref{sec:analysis}, we define and prove some important properties of matching algorithms in double sided auctions. In particular we present a dynamic price matching algorithm which produces a maximum as well as a fair matching. In Section~\ref{sec:matchingInMarkets}, we describe a uniform price matching algorithm used for price discovery in financial markets. Moreover, we prove that it produces a matching which is maximal among all possible uniform matchings. We summarise the work in Section~\ref{sec:conclusion} with an overview of possible future works. The Coq formalization for notions and proofs in this paper is available at \cite{auctiongithub}.  

\section{Modeling double sided auctions}\label{sec:modeling}
An auction is a  competitive event, where goods and services are sold to the most competitive participants. The priority among participating traders is decided by various attributes of the bids and asks (e.g. price, time etc). This priority can be finally represented by ordering them in a sequence. Sequences are best represented using the list data structure in the Coq standard library \cite{standLib}.

\subsection{Bid, Ask and limit price}
In  any double sided auction multiple buyers and sellers place their orders to buy or sell a unit of an underlying product. The auctioneer matches these buy-sell requests  based on their \emph{limit prices}. While the limit price for a buy order (i.e. \emph{bid}) is the price above which the buyer does not want to buy the item, the limit price of a sell order (i.e. \emph{ask}) is the price below which the seller does not want to sell the item. If a trader wishes to buy or sell multiple units, he can create multiple bids or asks with different \emph{ids}. 

We can express bids as well asks using  records containing two fields. 

\begin{verbatim}
  Record Bid: Type := Mk_bid { bp:> nat;    idb: nat }.
  Record Ask: Type := Mk_ask { sp:> nat;    ida: nat }.
\end{verbatim}

For a bid $b$, $(bp \; b)$  is the limit price and $(idb \; b)$ is its unique identifier. Similarly for an ask $a$, $(sp \; a)$ is the limit price and $(ida \; a)$ is the unique identifier of $a$. Note that the limit prices are natural numbers when expressed in the monetary unit of the lowest denomination (like cents in USA). Also note the use of coercion symbol \tw{ :>} in the first field of $Bid$. It declares $bp$ as an implicit function which is applied to any term of type $Bid$ appearing in a context where a natural number is expected. Hence from now on we can simply use $b$ instead of ($bp \;b$) to express the limit price of $b$. Similarly we can use $a$ for the limit price of an ask $a$.

Since equality for both the fields of $Bid$ as well as $Ask$ is decidable (i.e. \tw{nat: eqType}), the equality on $Bid$ as well as $Ask$ can also be proved decidable. This is achieved by declaring two canonical instances \tw{bid\_eqType} and \tw{ask\_eqType} which connect $Bid$ and $Ask$ to the \tw{eqType}.  

\subsection{Matching in Double Sided Auctions}
All the buy and sell requests can be assumed to be present in list $B$ and list $A$ respectively. At the time of auction, the auctioneer matches bids in $B$ to asks in $A$. We say a bid-ask pair $(b, a)$ is \emph{matchable} if $b \ge a$ (i.e. $bp \; b \ge sp \; a$).  Furthermore, the auctioneer assigns a trade price to each matched bid-ask pair. This  process results in  a matching $M$, which consists of all the matched bid-ask pairs together with their trade prices. We define matching as a list whose entries are of type \tw{fill\_type}.
\begin{verbatim}
  Record fill_type: Type:=  Mk_fill {bid_of: Bid;  ask_of: Ask;  tp: nat} 
\end{verbatim}

\begin{figure}[h!]
\centering
\includegraphics[width=.6\textwidth]{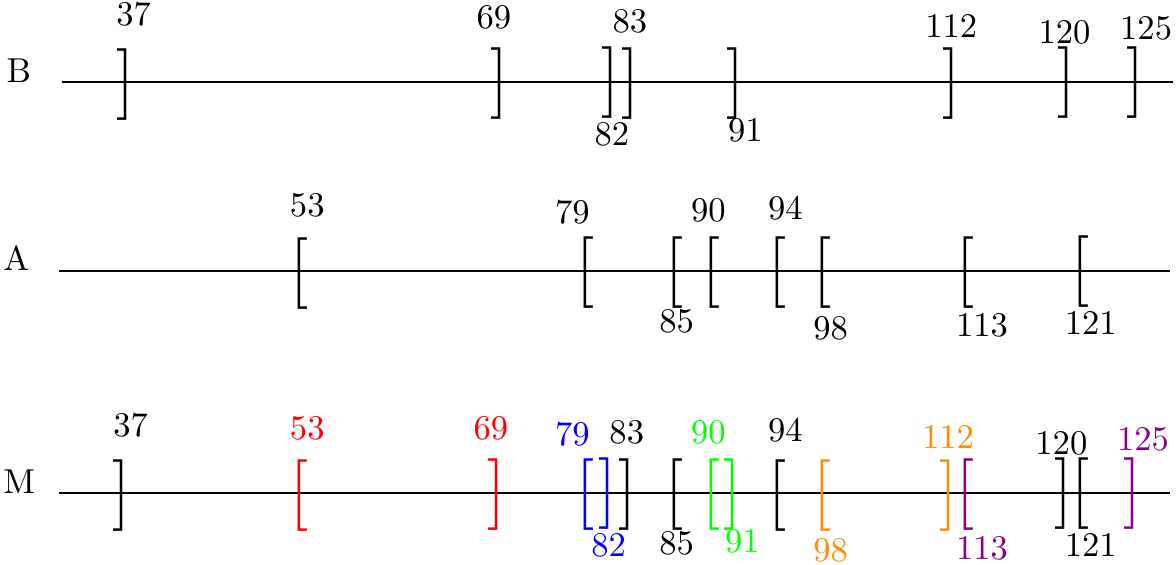}
\caption{ Bids in $B$ and asks in $A$ are represented using right and left brackets respectively along with their limit prices. Every matched bid-ask pair in $M$ is shown using  brackets of same colors. Bids with limit prices $37$, $83$ and $120$ are not matched to any ask in the matching $M$.}
\label{fig:matching}
\end{figure}

In a matching $M$, a bid or an ask appears at most once. Note that there might be some bids in $B$ which are not matched to any asks in $M$. Similarly there might be some asks in $A$ which are not matched to any bids in $M$. The list of bids present in $M$ is denoted by $B_{M}$ and the list of asks present in $M$ is denoted by $A_M$. For example, Fig.~\ref{fig:matching} shows a matching $M$ between list of bids $B$ and list of asks $A$. Note that the bid with limit price $37$ is not present in $B_M$ since it is not matched to any ask in $M$.  

More precisely, for a given list of bids $B$ and list of asks $A$, $M$ is a matching iff, (1) All the bid-ask pairs in $M$ are matchable, (2) $B_M$ is duplicate-free, (3) $A_M$ is duplicate-free, (4) $B_M \subseteq B$, and (5) $A_M \subseteq A$.

\begin{figure}[h!]
\centering
\includegraphics[width=.5\textwidth]{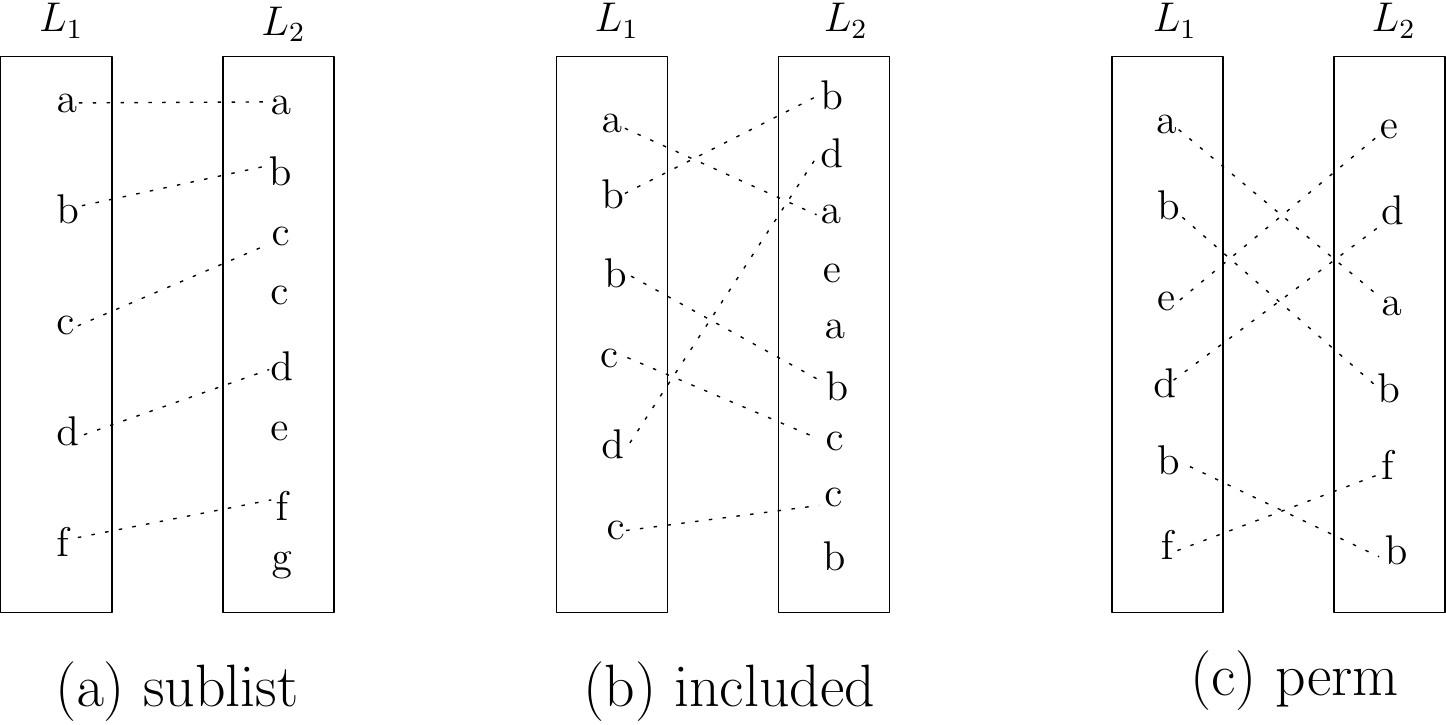}
\caption{The dotted lines between the entries of lists confirm the presence of these entries in both the lists. (a) If $L_1$ is \tw{sublist} of $L_2$ then every entry of $L_1$ is also present in $L_2$ and they appear in the same succession. (b) A list $L_1$ is \tw{included} in $L_2$ if every entry in $L_1$ is also present in $L_2$. (c) Two lists $L_1$ and $L_2$ are permutation of each other if each entry has same number of occurrences in both  $L_1$ and $L_2$. }
\label{fig:list}
\end{figure}

\begin{definition}
\tw{ \textcolor{gray}{matching\_in} B A M := All\_matchable M $\land$ NoDup $B_M$ $\land$ NoDup $A_M$ $\land$ $B_M \subseteq B$ $\land$ $A_M \subseteq A$.}
\end{definition}
The term \emph{\tw{NoDup $B_M$}} in  the above definition indicates that each bid is a request to trade one unit of item and the items are indivisible.  We use the expression $B_M \subseteq B$  to denote the term \tw{(Subset $B_M$ $B$)}.  It expresses the fact that each element in the list $B_M$ is also present in the list $B$. While the predicates \tw{NoDup} and  \tw{Subset}  are sufficient to express the notion of a matching, we need more definitions to describe the properties of matching in double sided auctions.  In Fig \ref{fig:list} we describe three binary relations on lists namely \tw{sublist}, \tw{included} and \tw{perm} which are useful in stating some intermediate lemmas leading to important results on matching. For example, consider the following lemma which states that the property of being a matching is invariant under permutation. 

\begin{lemma}
\tw{\textcolor{gray}{match\_inv}: perm M  M' -> perm B B' -> perm A A' -> matching\_in B A M -> matching\_in B' A' M'. }
\end{lemma}

Note that the notion of permutation for lists is analogous to the equality in multisets. More precisely, we have the following lemmas specifying the \tw{perm} relation.

\begin{lemma}\label{lem:permCount} 
\tw{ \textcolor{gray}{ perm\_intro}: ($\forall$ a, count a l = count a s) -> perm l s.}
\end{lemma}
\begin{lemma}
\tw{ \textcolor{gray}{perm\_elim}: perm l s -> ($\forall$ a, count a l = count a s).}
\end{lemma}

The term (\emph{\tw{count a l}}) in Lemma~\ref{lem:permCount}  represents the number of occurrences of element $a$ in the list $l$. In proving various properties of a matching $M$ we very often base our arguments solely on the information present in $B_M$, $A_M$ and $P_M$. Therefore it is useful to have lemmas establishing the interaction of  $B_M$, $A_M$ and $P_M$ with above mentioned relations on lists. 

\begin{lemma}
\tw{\textcolor{gray}{included\_M\_imp\_included\_bids}:included $M$  $M'$ -> included $B_M$ $B_{M'}$ }.
\end{lemma}
\begin{lemma}
\tw{\textcolor{gray}{included\_M\_imp\_included\_asks}:included $M$  $M'$ -> included $A_M$ $A_{M'}$ }
\end{lemma}

The notion of included in above lemmas is similar to subset relation on multisets. We have the following lemmas specifying the exact behaviour of \tw{included} relation.
\begin{lemma}
\tw{\textcolor{gray}{included\_intro}: ($\forall$ a, count a l $\leq$ count a s)-> included l s.}
\end{lemma}
\begin{lemma}
\tw{ \textcolor{gray}{included\_elim}: included l s -> ($\forall$ a, count a l $\leq$ count a s).}
\end{lemma}

In this work, we come across various processes whose input and output are lists. We need the \tw{sublist} relation, which is similar to the subsequence relation, to properly specify the behaviour of these processes.  

\begin{lemma} 
\tw{ \textcolor{gray}{sublist\_intro1} (a:T): sublist l s-> sublist l (a::s).}
\end{lemma}
\begin{lemma}\label{lem:sublistInd} 
\tw{ \textcolor{gray}{sublist\_elim3a} (a e:T): sublist (a::l)(e::s)-> sublist l s.}
\end{lemma}

Note the recursive nature of \tw{sublist} as evident in Lemma~\ref{lem:sublistInd}. It  makes inductive reasoning easier for the statements which contain \tw{sublist} in the antecedent. However, this is  not true for the other relations (i.e. \tw{included} and \tw{perm}).

\section{Analysis of Double sided auctions}\label{sec:analysis}
\begin{figure}[h!]
\centering
\includegraphics[width=.6\textwidth]{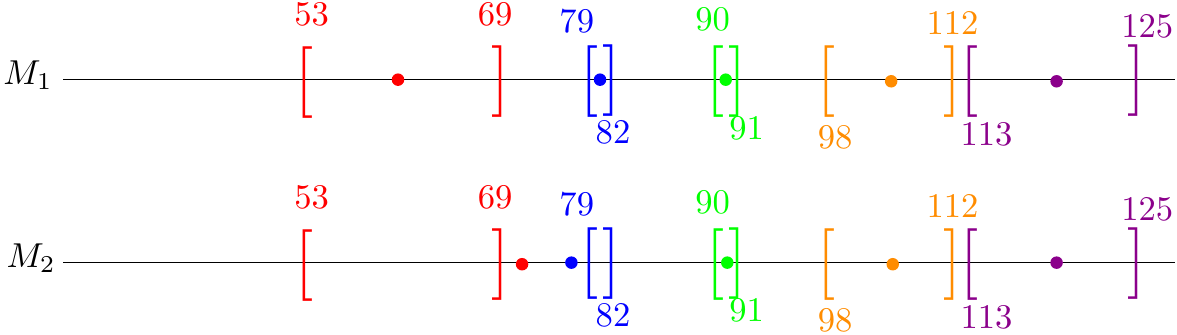}
\caption{ The colored dots represent the trade prices at which the corresponding matched bid-ask pairs are traded. While the matching $M_2$ is not IR since some dots lie outside the corresponding matched bid ask-pair. The matching $M_1$ is IR because trade prices for every matched bid-ask pair  lie inside the interval. Note that the matching $M_1$ and $M_2$ contains exactly the same bid-ask pairs.}
\label{fig:IR}
\end{figure}

In this, work we do not consider analysis of profit for the auctioneer. Therefore the buyer of a matched bid-ask pair pays the same amount which the seller receives. This price for a matched  bid-ask pair is called the trade price for that pair. Since the limit price for a buyer is the price above which she does not want to buy, the trade price for this buyer is expected to be below its limit price. Similarly the trade price for the seller is expected to be above its limit price. Therefore in any matching it is desired that the trade price of a bid-ask pair lies between their limit prices. A matching which has this property is called an \emph{individual rational (IR)} matching. Note that any matching can be converted to an IR matching without altering its bid-ask pair (See Fig~\ref{fig:IR}).

The number of matched bid-ask pairs produced by a matching algorithm is crucial in the design of a double sided auction mechanism. Increasing the number of matched bid-ask pairs increases  liquidity in the market. Therefore, producing a maximum matching is an important aspect of double sided auction mechanism design. For a given list of bids $B$ and list of asks $A$ we say a matching $M$ is  a maximum matching if no other matching $M'$ on the same $B$ and $A$ contains more matched bid-ask pairs than $M$. 

\begin{definition}
\tw{ \textcolor{gray}{Is\_MM} M B A := (matching\_in B A M) $\land$ 
($\forall$ M', matching\_in B A M' $\rightarrow$ |M'| $\leq$ |M|)}.
\end{definition}

\begin{figure}[h!]
\centering
\includegraphics[width=.35\textwidth]{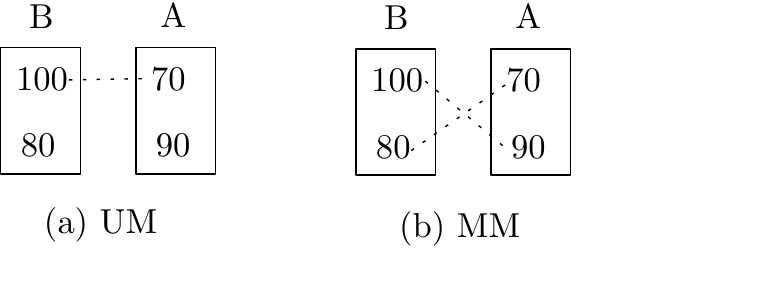}
\caption{In this case the only individually rational matching of size two it is not uniform.}
\label{fig:mmum}
\end{figure}

In certain situations, to produce a maximum matching, different bid-ask pairs must be assigned different trade prices (Fig ~\ref{fig:mmum}).  However, different prices for the same product in the same market simultaneously leads to dissatisfaction amongst some of the traders. A mechanism which clears all the matched bid-ask pairs at same trade price is called a \emph{uniform matching}. It is also known as \emph{perceived-fairness}. 

\subsection{Fairness}  

A bid with higher limit price is more competitive compared to bids with lower limit prices. Similarly an ask with lower limit price is more competitive compared to asks with higher limit prices. In a competitive market more competitive traders are prioritised for matching. A matching which prioritises more competitive traders is called a \emph{fair} matching. 

\begin{definition}
\tw{ \textcolor{gray}{fair\_on\_bids} M B:=
$\forall$ b b', In b B $\land$ In b' B -> b $>$ b' -> In b' $B_M$ -> In b $B_M$.}
 \end{definition}
\begin{definition}
\tw{ \textcolor{gray}{fair\_on\_asks} M A:=
$\forall$ s s', In s A $\land$ In s' A -> s $<$ s' -> In s' $A_M$ -> In s $A_M$. }
\end{definition}
\begin{definition}
\tw{ \textcolor{gray}{Is\_fair} M B A:=
fair\_on\_asks M A $\land$ fair\_on\_bids M B.}
\end{definition}

Here, the predicate \tw{fair\_on\_bids M B}  states that the matching $M$  is fair for the list of buyers $B$. Similarly,  the predicate \tw{fair\_on\_asks M A} states that the matching $M$  is fair for the list of sellers $A$.  A matching which is fair on bids as well as ask is expressed using the predicate \tw{Is\_fair M B A}.

Unlike the uniform matching, a fair matching can always be achieved without compromising the size of matching. We can accomplish this by converting any matching into a fair matching without changing its size. For example, consider the following function \tw{make\_FOB}.

\begin{verbatim}
	  Fixpoint Make_FOB (M:list fill_type) (B: list Bid):=
	    match (M,B) with 
	    |(nil,_) => nil
	    |(m::M',nil) => nil
	    |(m::M',b::B') => (Mk_fill b (ask_of m) (tp m))::(Make_FOB M' B')
	  end.
\end{verbatim}

\begin{figure}[h!]
\centering
\includegraphics[width=.6\textwidth]{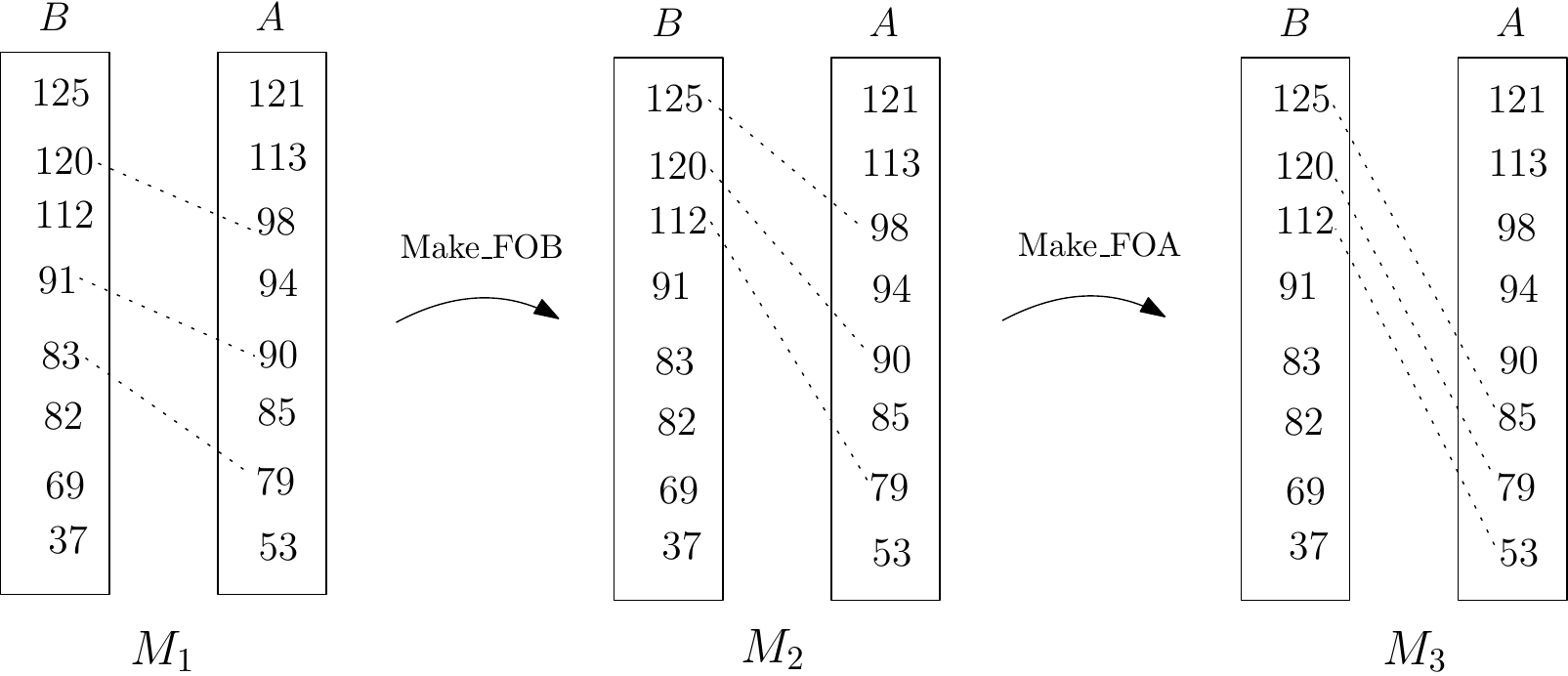}
\caption{The dotted lines in this figure represent  matched bid-ask pairs in  matching $M_1$, $M_2$ and $M_3$. In the first step function \tw{make\_FOB} operates on $M_1$ recursively. At each step it picks the top bid-ask pair, say $(b,a)$ in $M_1$ and replaces the bid  $b$ with a most competitive bid available in $B$. The result of this process is a \tw{fair\_on\_bids} matching $M_2$. In a similar way the function \tw{make\_FOA} changes $M_2$ intro a fair on ask matching $M_3$.  }
\label{fig:fair}
\end{figure}

The function \tw{make\_FOB} produces a \tw{fair\_on\_bids} matching from a given matching M and a list of bids B, both sorted in decreasing order of bid prices (See Fig~\ref{fig:fair}).  Note that \tw{make\_FOB} doesn't change any of the ask in M and due to the recursive nature of \tw{make\_FOB} on B, a bid is not repeated in the process of replacement.  Hence the new $B_M$ is duplicate-free. Once we get a fair matching on bids, we use similar function \tw{make\_FOA} to produce a fair matching. 

More precisely, for the function \tw{make\_FOB} and \tw{make\_FOA} we have the following lemmas proving it fair on bids and fair on asks respectively. 
\begin{lemma}\label{lem:fob}
\tw{ \textcolor{gray}{mfob\_fair\_on\_bid} M B:
  (Sorted M) -> (Sorted B) -> sublist $P_{B_M}$ $P_B$ ->
  fair\_on\_bids (Make\_FOB M B) B.  }
\end{lemma}

\begin{lemma}\label{lem:foa}
\tw{ \textcolor{gray}{mfob\_fair\_on\_ask} M A: 
 (Sorted M) -> (Sorted A) -> sublist $P_{A_M}$ $P_A$ ->
  fair\_on\_asks (Make\_FOA M A) A.  }
\end{lemma}

\begin{theorem}\label{thm:fairMatching}
\tw{ \textcolor{gray}{exists\_fair\_matching} (Nb: NoDup B)(Na: NoDup A): 
 matching\_in B A M -> ( $\exists$ M', matching\_in B A M' $\land$ Is\_fair M' B A $\land$ |M| = |M'|). }
\end{theorem}

Proof of Theorem~\ref{thm:fairMatching} depends on Lemma \ref{lem:foa} and Lemma \ref{lem:fob}. Furthermore, Lemma \ref{lem:foa} and Lemma \ref{lem:fob} can be proved using induction on the size of M. 

\subsection{Maximum Matching}

\begin{figure}[h!]
\centering
\includegraphics[width=.6\textwidth]{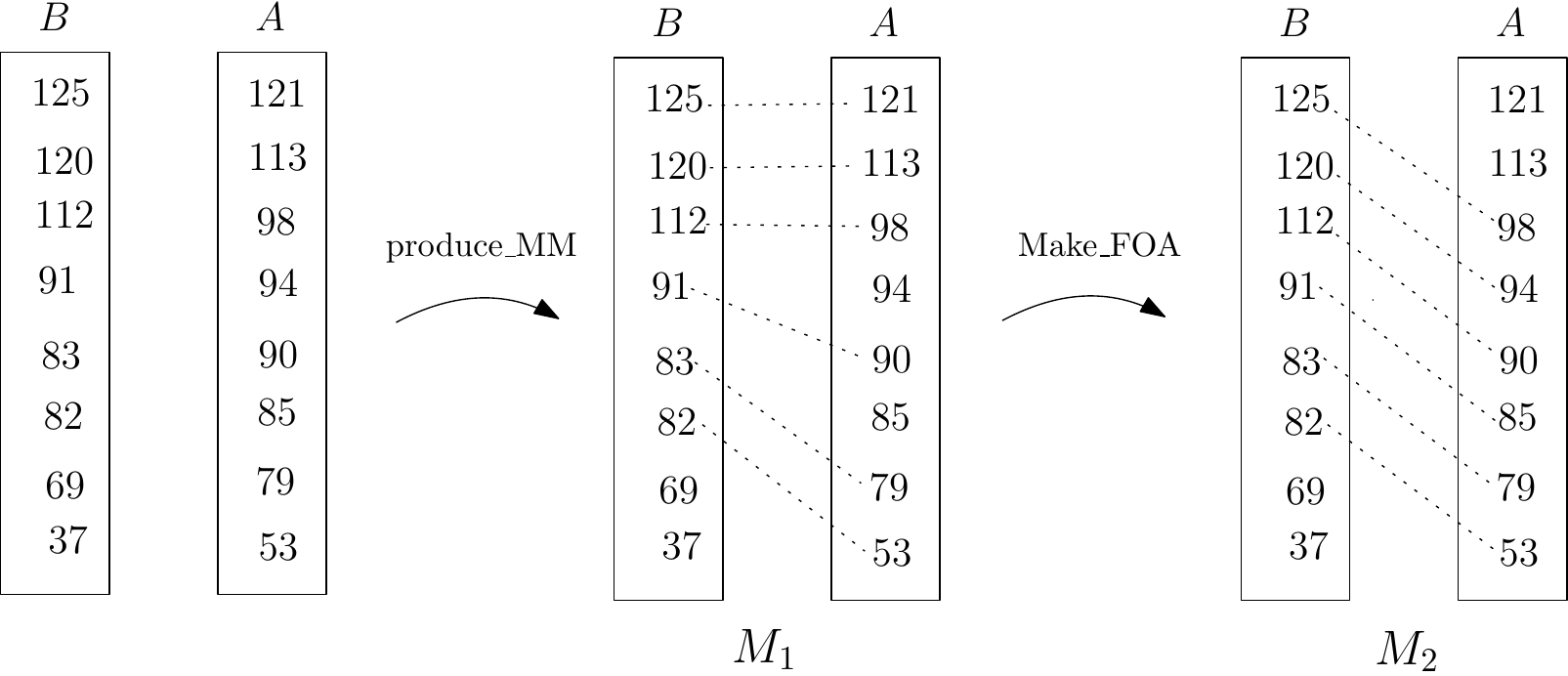}
\caption{In the first step, the function \tw{produce\_MM} operates recursively on the list of bids B and list of asks A. At each iteration \tw{produce\_MM} selects a most competitive available bid and then pairs it with the largest matchable ask. The output of this function is fair on bids since it doesn't leave any bid from top. In the second step, the function \tw{make\_FOA} converts  $M_1$ into fair matching $M_2$. }
\label{fig:mm}
\end{figure}

The liquidity in any market is a measure of how quickly one can trade in the market without much cost. One way to increase the liquidity is to maximize the number of matched bid-ask pairs. In the previous section we have seen that any matching can be changed to a fair matching without altering its size. Therefore, we can have a maximum matching without compromising on the fairness of the matching. In this section we describe a matching which is fair as well as maximal. For a given list of bid $B$ and list of ask $A$, a maximum and fair matching can be achieved in two steps. In the first step we apply function \tw{produce\_MM} which produces a matching which is maximal and fair on bids. In the next step we apply \tw{make\_FOA} to this maximum matching to produce a fair matching (See Fig~\ref{fig:mm}).

\begin{verbatim}
Fixpoint produce_MM (B:list Bid) (A: list Ask): (list fill_type) :=
  match (B, A) with
  |(nil, _) => nil
  |(b::B', nil) => nil              
  |(b::B', a::A') =>  match (a <= b) with
     |true => ({|bid_of:= b; ask_of:= a; tp:=(bp b)|})::(produce_MM B' A')
     |false => produce_MM B A'
    end
  end. 
\end{verbatim}

At each iteration  \tw{produce\_MM} generates a matchable bid-ask pair (See Fig~\ref{fig:mm}). Due to the recursive nature of  function  \tw{produce\_MM} on both $B$ and $A$, it never pairs any bid with more than one asks. This ensures that the list of bids in matching (i.e. $B_M$) is duplicate-free. Note that  \tw{produce\_MM} tries to match a bid until it finds a matchable ask. The function terminates when either all the bids are matched or it encounters a bid for which no matchable ask is available.  Therefore, the function \tw{produce\_MM} produces a matching from a given lists of bids $B$ and a list of asks $A$, both sorted in decreasing order by there limit prices.  The following theorem states that the function \tw{produce\_MM} produces a maximum matching when both  $B$ and $A$ are sorted in decreasing order by limit prices.
\begin{theorem}
\tw{\textcolor{gray}{produce\_MM\_is\_MM}(Nb:NoDup B)(Na:NoDup A): Sorted  B -> Sorted A -> Is\_MM (produce\_MM B A) B A. }
\end{theorem}
\textbf{Proof}: We prove this result using induction on the size of list $A$. 
\begin{itemize}
\item Induction hypothesis (\tw{IH}): \emph{ \tw{ $\forall$ A', |A'| < |A| -> $\forall$ B, Sorted  B -> Sorted A' -> Is\_MM (produce\_MM B A') B A'.} }
\end{itemize}
Let $M$ be an arbitrary matching on the list of bids $B$ and list of asks $A$. Moreover, assume that $b$ and $a$ are the topmost bid and ask present in $B$ and $A$ respectively (i.e. $A = (a::A')$ and $B = (b::B')$). We prove $|M| \leq$ |\tw{produce\_MM} $B$ $A$| in  the following two cases.
\begin{itemize}
\item \textbf{Case-1} ($b<a$): In this case the limit price of $a$ is strictly more than the limit price of $b$.  In this case the function \tw{produce\_MM} computes a matching on $B$ and $A'$. Note that due to the induction hypotheses (i.e. \tw{IH}) this is a maximum matching for $B$ and $A'$. Since the limit price of ask $a$ is more than the most competitive bid $b$ in $B$ it cannot be present in any matching of $B$ and $A$. Therefore a maximum matching on $B$ and $A'$ is also a maximum matching on $B$ and $A$. Hence we have |$M$| $\leq$ | \tw{produce\_MM} $B$ $A$ |.
\item \textbf{Case-2} ($a \leq b$): In this case \tw{produce\_MM} produces a matching of size $m + 1$ where $m$ is the size of matching \tw{produce\_MM $B'$ $A'$}. We need to prove that \tw{$|M|$ $\leq$ $m+1$}. Note that due to induction hypothesis (i.e. \tw{IH}) the matching \tw{produce\_MM $B'$ $A'$} is a maximum matching on $B'$ and $A'$.  Hence no matching on $B'$ and $A'$ can have size bigger than $m$. Without loss of generality we can assume that $M$ is also sorted in decreasing order of bid prices. Now we further split this case into the following five sub cases (see Fig~\ref{fig:mmProof}).

\begin{figure}[h!]
\centering
\includegraphics[width=0.7\textwidth]{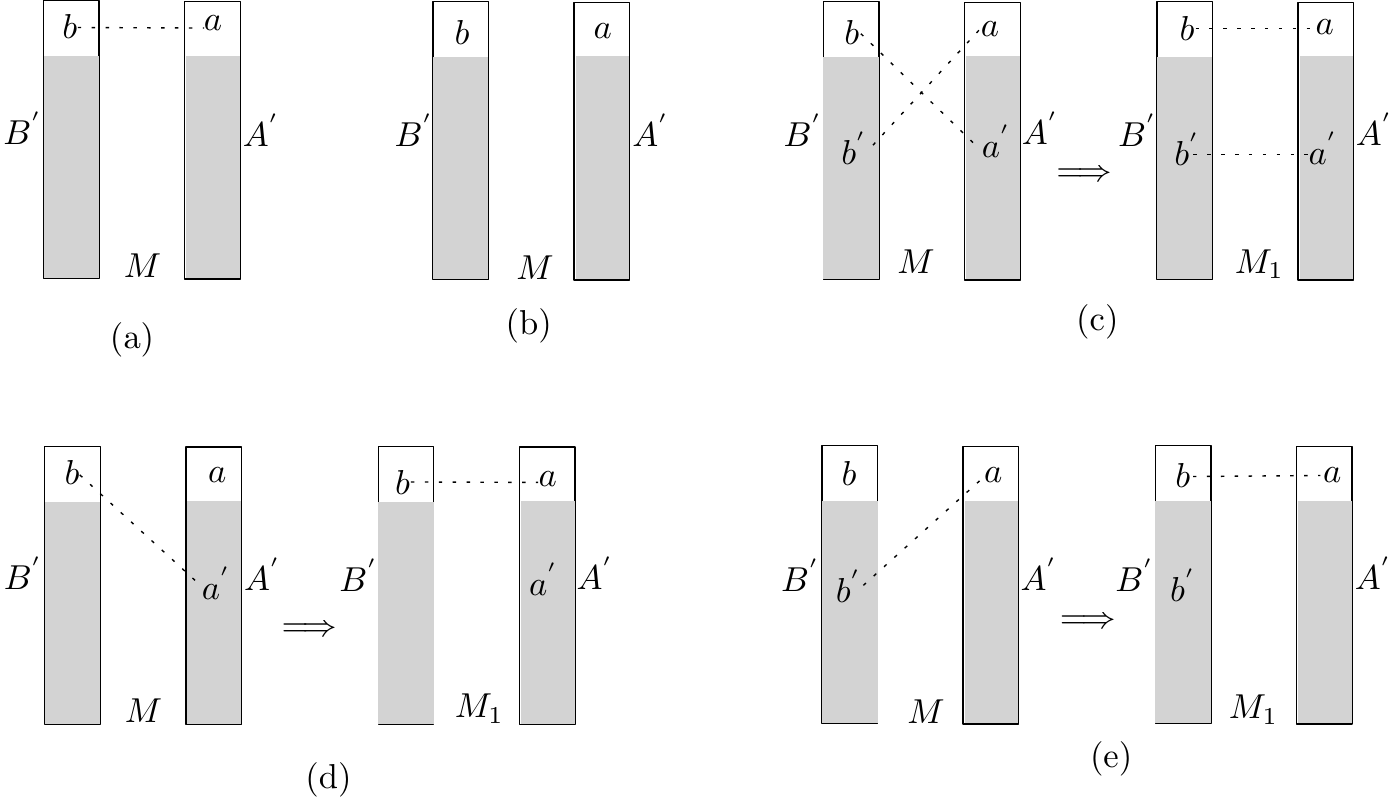}
\caption{This figure shows all the five sub cases of Case-2 (i.e. when $b \geq a$). The dotted line shows presence of the connected pair in matching $M$.  Both the list of bids $B$ and list of asks $A$ are sorted in decreasing order of their limit prices. Moreover, we assume $B = b::B'$ and $A = a::A'$.  }
\label{fig:mmProof}
\end{figure}

\begin{itemize}
\item \textbf{Case-2A} ($M = (b,a)::M'$) : In this case  bid $b$ is matched to ask $a$ in the matching $M$ (see Fig~\ref{fig:mmProof} (a)). Note that $M'$ is a matching on $B'$ and $A'$. Since $|M'| \leq m$ we have $|M|= |M'|+1 \leq m+1$.
\item \textbf{Case-2B} (\tw{$b \notin B_M  \land a \notin A_M$}) : In this case neither bid $b$ nor ask $a$ is present in matching $M$ (see Fig~\ref{fig:mmProof} (b)). Therefore $M$ is a matching on $B'$ and $A'$. Hence we have $|M| \leq m < m+1$.
\item \textbf{Case-2C} $(b,a') \in M \land (b',a) \in M$ : In this case we have $(b,a') \in M$ and  $(b',a) \in M$ where $a' \in A'$ and $b' \in B'$. We can obtain another matching $M_1$ of same size as $M$ (see Fig~\ref{fig:mmProof} (c)) where $(b,a) \in M_1$ and $(b',a') \in M_1$. Note that all other entries of $M_1$ is same as $M$. Therefore we have $M_1 = (b,a)::M'$ where $M'$ is a matching on $B'$ and $A'$. Since $|M'| \leq m$ we have $|M|=|M_1| \leq m+1$.
\item \textbf{Case-2D}: $(b,a')\in M \land a \notin A_M$ : In this case we have $(b,a')\in M$ and $a \notin A_M$ where $a'\in A'$. We can obtain another matching $M_1$ of same size as $M$ (see Fig~\ref{fig:mmProof} (d)) where $(b,a) \in M_1$. 
Therefore we have $M_1 = (b,a)::M'$ where $M'$ is a matching on $B'$ and $A'$.
Since $|M'| \leq m$ we have $|M|=|M_1| \leq m+1$.
\item \textbf{Case-2E}: $(b',a)\in M \land b \notin B_M$ : In this case we have $(b',a)\in M$ and $b \notin B_M$ where $b'\in B'$. We can obtain another matching $M_1$ of same size as $M$ (see Fig~\ref{fig:mmProof} (e)) where $(b,a) \in M_1$. 
Therefore we have $M_1 = (b,a)::M'$ where $M'$ is a matching on $B'$ and $A'$.
Since $|M'| \leq m$ we have $|M|=|M_1| \leq m+1$.
\end{itemize}
\end{itemize}
Note that all the cases in the above proof correspond to  predicates which can be expressed using  only the membership predicate on lists. Since we have  decidable equality on the elements of the lists all these predicates are also decidable. Hence, we can do case analysis on them without assuming any axiom. $\square$

Now that we  proved the maximality property of \tw{produce\_MM}  we can produce a fair as well as maximal matching by applying the functions \tw{Make\_FOA}  and \tw{Make\_FOB} to the output of \tw{produce\_MM}. More precisely, for a given list of bids $B$ and list of asks $A$, we have  following result stating that there exists a matching which is both maximal and fair. 

\begin{theorem}
\tw{\textcolor{gray}{exists\_fair\_maximum} (B: list Bid)(A: list Ask): $\exists$ M, (Is\_fair M B A $\land$ Is\_MM M B A).}
\end{theorem}

\subsection{Matching in financial markets}\label{sec:matchingInMarkets} 

An important aspect of the pre-market session is to discover a single price (equilibrium price) at which maximum demand and supply can be matched. Most exchanges execute trade during this session at an equilibrium price.  Consider the function  \tw{UM} which produces an individually rational matching which is fair and maximal among all uniform matchings.

\begin{verbatim}
Fixpoint produce_UM (B:list Bid) (A:list Ask)  :=
  match (B,A) with
  |(nil, _) => nil
  |(_,nil)=> nil
  |(b::B',a::A') =>  match (a <= b) with
     |false =>nil
     |true  => ({|bid_of:= b; ask_of:= a; tp:=(bp b)|})::produce_UM B' A'
  end
end.
Definition uniform_price B A := bp (bid_of (last (produce_UM B A))).
Definition UM B A:= replace_column (produce_UM B A) (uniform_price B A).
\end{verbatim}

\begin{figure}[h!]
\centering
\includegraphics[width=0.35\textwidth]{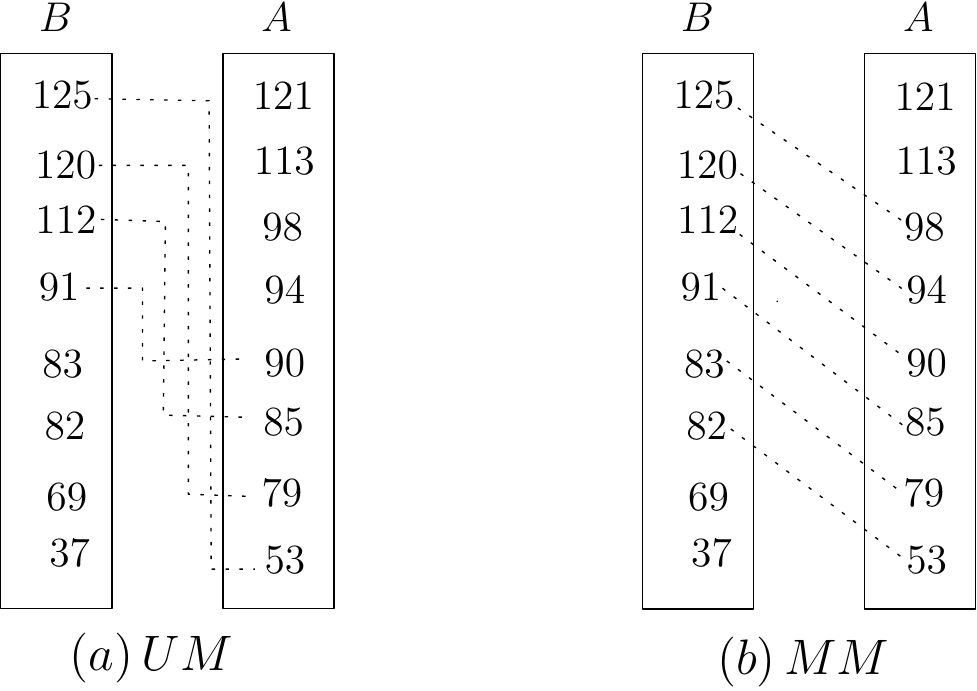}
\caption{(a) The dotted lines indicate all the bid-ask pair  produced by function \tw{produce\_UM}. In each iteration function \tw{produce\_UM}  matches the largest available bid in $B$ with the smallest available ask in $A$.  (b) The dotted lines here indicate a maximum matching for the list of bids $B$ and list of asks $A$. Note that in this case the matching produced by \tw{UM} is not a maximum matching. }
\label{fig:UM}
\end{figure}

The function \tw{produce\_UM} produces bid-ask pairs, \tw{uniform\_price} computes the uniform price and finally \tw{UM} produces a uniform matching. The function \tw{produce\_UM} is recursive and  matches the largest available bid in $B$ with the smallest available ask in $A$ at each iteration (See Fig~\ref{fig:UM}). This function terminates when the most competitive bid available in $B$ is not matchable with any available ask in $A$. The following theorem states that the function \tw{produce\_UM} produces a maximal matching among all uniform matchings when the list of bids $B$ is sorted in decreasing order by limit prices and the list of asks $A$ is sorted in increasing order by limit prices.

\begin{theorem}
\tw{
\textcolor{gray}{UM\_is\_maximal\_Uniform} (B: list Bid) (A:list Ask): Sorted  B -> Sorted A -> $\forall$ M: list fill\_type, Is\_uniform M -> |M| $\leq$ | (UM B A ) |
}
\end{theorem}

\begin{figure}[h!]
\centering
\includegraphics[width=0.6\textwidth]{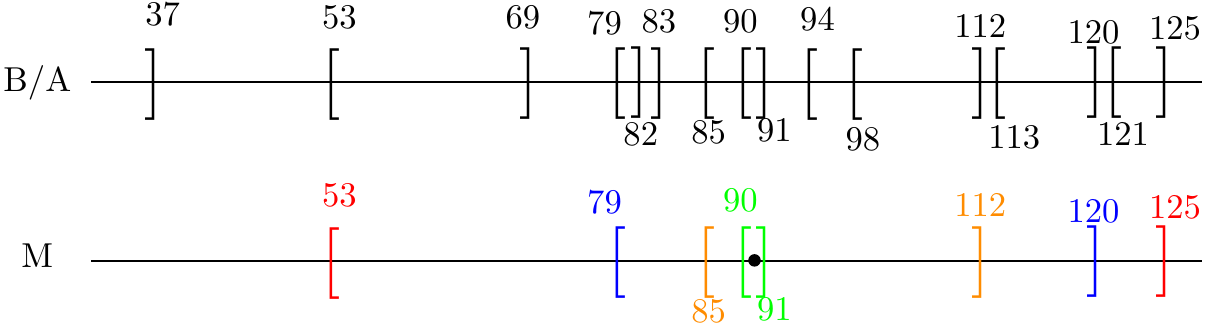}
\caption{ Trade price $p$ for the matching $M$ is shown using a dot that lies between the ask with limit price 90 and bid with limit price 91. Note that since $M$ is individually rational the number of matched asks below the trade price $p$ is same as number of matched bids above the trade price $p$. }
\label{fig:UM_match}
\end{figure}

\textbf{Proof}: Let $M$ be any arbitrary IR and uniform matching on the list of bids $B$ and list of asks $A$ where each matched bid-ask pair is traded at price $t$. We need to prove that $m \leq$ |\tw{(UM $B$ $A$)}| where $m$ is the number of matched bid-ask pairs in the matching $M$. Observe that in any individually rational and uniform matching the number of bids above  the trade price is same as the number of asks below the trade price (See Fig~\ref{fig:UM_match}). Therefore, there are at least $m$ bids above $t$ and $m$ asks below $t$ in $B$ and $A$ respectively. 

Since at each step the function \tw{produce\_UM}  pairs the largest bid available in $B$ with the smallest ask available in $A$ it must produce  at least $m$ bid-ask pairs. Hence for the list of bids $B$ and list of asks $A$ the function \tw{UM} produces a uniform matching which is of size at least $m$. $\square$

\section{Conclusion}\label{sec:conclusion}
Trading activities in today's financial markets are mostly enabled using computer algorithms. These algorithms are extremely large and complex. Matching algorithms used by exchanges (venues) are at the core of this broad range of financial algorithms \cite{PassmoreI17}. To ensure safety and integrity in the markets, the market regulators introduce guidelines specifying different features for these algorithms. Traditional methods of software development, which focus on testing, can not guarantee that these softwares meet the guidelines. 

In this work, we develop a formal framework to verify some important properties of the matching algorithms used by exchanges. These algorithms use double sided auctions to match multiple buyers with multiple sellers during different sessions of trading. We use the dependent types of Coq proof assistant to concisely represent various notions from auction theory relevant for the verification of these algorithms. We formally verify two important classes of double sided auctions (uniform price and dynamic price) in this framework. 

In this work, we define each bid or ask as a request to trade a single unit of a product and the product is indivisible. In the future this work can be extended to accommodate trades involving multiple units of an item by introducing proper functions to generate bids and asks of single unit from the buy and sell requests of multiple units. Another interesting direction of work is to extend this work for different types of orders (e.g. limit orders, market orders, stop-loss orders, iceberg orders etc) in continuous markets. It would require maintaining a priority queue based on the various attributes of these orders. A formal verification of trading at an exchange will provide a formal foundation that can be used for rigorous analysis of other financial algorithms (e.g. order routing, clearing and settlements etc).

\bibliography{auction}

\end{document}